\documentclass[12pt]{article}
\begin{document}

\centerline{\bf ELIMINATION OF SPURIOUS STATES OF OSCILLATOR SHELL MODEL}
\centerline{A.Deveikis and G.Kamuntavi{\v c}ius}
\centerline{Vytautas Magnus University, 3000 Kaunas, Lithuania}
\date{\today}

\begin{abstract}
A new method for elimination of the 'spurious' states comprising requirement
of translational invariance and simplicity of the enumeration scheme of
antisymmetric $A$-particle states has been developed.
The method presented enables one to project off the space of purely spurious
antisymmetric $A$-particle oscillator functions with singled out dependence
on intrinsic coordinates of two last particles onto the subspace of the
nonspurious motion.
The method is based on constructing the matrix elements of the centre-of-mass
Hamiltonian and spectral decomposition of the corresponding real
symmetric matrix.
The obtained coefficients of the expansion of the antisymmetric $A$-particle
oscillator functions in terms of the ones with singled out dependence on intrinsic
coordinates of two last particles and with eliminated spurious states are
required for calculation of two-particle oscillator intrinsic density matrices.
Software tools based on a root rational fraction expression of numbers for the
theoretical formulation presented have been developed.
\end{abstract}

\section{Introduction}

The translationally invariant shell model, originally proposed by
Kretzschmar \cite{Kretzschmar1,Kretzschmar2}, offers a very sophisticated
method for characterization of antisymmetric states and producing the
intrinsic coefficients of fractional parentage (ICFP's).
A description of the many-particle wave functions
in terms of relative coordinates turns out to be unfeasible because of the
antisymmetrization requirements of the Pauli principle.
As a consequence the ICFP's are available only in the E$_{\min }$ approach
and for the excited states with the maximum orbital momentum
\cite{Vanagas71,Shitikova70}.

More promising is to establish and use the relation of the translationally
invariant shell model (for an arbitrary excitation) and the ordinary oscillator
shell model.
The relation of this type makes it possible to obtain the results
that obey the principle of translational invariancy by means of calculations
within the framework of considerably simpler oscillator shell model.

However, the ordinary oscillator shell model is inherent with the so-called
'spurious' states corresponding to the centre-of-mass motion of the nucleus
\cite{Wildermuth77}.
The ordinary oscillator shell model Hamiltonian is not invariant with respect
to translations.
Hence, the wave functions under calculation may contain nonphysical
components, which are called spurious.
The latter describe excitations of the centre-of-mass motion of the
nucleus that are introduced with the single-particle potentials in the
laboratory system.
These excitations should be eliminated, since only the intrinsic
excitations corresponds to the genuine excitations of the physical system
that are observed.

The description of intrinsic properties of atomic nuclei may be simplified
using the two-particle intrinsic density matrices instead of wave functions
\cite{LF96}.
However, according the procedure proposed in Ref. \cite{LF96}, for calculation of
the two-particle intrinsic density matrix for an arbitrary number of
oscillator quanta, the expansion coefficients for the product of the
centre-of-mass ground state function and the intrinsic wave function in
terms of antisymmetric $A$-particle functions are needed.

In the present work we show that the coefficients of this type may be
obtained by generating purely spurious antisymmetric $A$-particle oscillator
functions with singled out dependence on intrinsic coordinates of two last
particles and, then, by projecting them off onto the subspace of the
nonspurious motion.
This may be accomplished by diagonalizing the centre-of-mass Hamiltonian
in the basis set of antisymmetric $A$-particle oscillator functions with
singled out dependence on intrinsic coordinates of two last particles and
choosing the subspace of its eigenvectors corresponding to the minimal
eigenvalue equal to $\frac{3}{2}$.
This procedure can be successful only when the nontruncated configuration
space is taken into account.
It implies that a large number of antisymmetric $A$-particle states has to
be used.
Such large scale calculations can be carried out by using the algorithms
presented in Ref. \cite{LF96,LF95} for enumeration and formation of antisymmetric
$A$-particle states.
The proposed method for elimination of 'spurious' states is the first one
which enables practically production of two-particle oscillator intrinsic density
matrices for an arbitrary excitation.

\section{The centre-of-mass Hamiltonian matrix}

The functions from the subspace of the nonspurious motion are the
eigenfunctions of the centre-of-mass Hamiltonian.
Thus, projection onto the subspace of the nonspurious motion implies such
superposition of the oscillator shell model functions which diagonalize the
centre-of-mass Hamiltonian matrix.
So, at first the matrix elements of the centre-of-mass Hamiltonian
between the osillator shell model functions (in the configuration space) should
be obtained, at second, the eigenvalue problem should be solved, and at last
the submatrix corresponding to the zero excitation energy should be extracted.

The Hamiltonian of the many-particle oscillator shell model may be divided in
two parts: the
centre-of-mass Hamiltonian $H_{\mbox{\small c.m.\vphantom{intr}}}$ and
the intrinsic Hamiltonian $H_{\mbox{\small intr.}}$

\abovedisplayskip=0pt

\begin{eqnarray}
H = H_{\mbox{\small c.m.\vphantom{intr}}} + H_{\mbox{\small intr.}}\ .
\end{eqnarray}

The centre-of-mass  Hamiltonian $H_{\mbox{\small c.m.\vphantom{intr}}}$
describes the motion of the centre of mass of the $A$-particle system
and is of the form

\begin{eqnarray} \label{Hmc_viend0}
H_{\mbox{\small c.m.\vphantom{intr}}}
= \textstyle\frac{\hbar\omega}{2}
\left\{ - \frac{b^2}{A}\vec\nabla_R^2 + {\frac{A}{b^2}} R^2 \right\}\ .
\end{eqnarray}

\noindent Here $\vec R=\frac{1}{A}\sum_{i=1}^A\vec r_i$ denotes the centre-of-mass
position vector, $\vec\nabla_R=\sum_{i=1}^A\vec \nabla_i$, and
$\omega$ is related with the usual oscillator length via
$b = \sqrt{\hbar/m\omega}$, where $m$ denotes the nucleon mass.
In the one-particle coordinates the Hamiltonian (\ref{Hmc_viend0}) can be
written as

\begin{eqnarray}  \label{Hmc_viend}
H_{\mbox{\small c.m.\vphantom{intr}}}
= \textstyle
\frac{\hbar\omega}{2A} \left\{ -b^2\left(
\displaystyle\sum_{i=1}^A\vec \nabla_i \right)^2 +
{\frac{1}{b^2}}\left(\displaystyle\sum_{i=1}^A\vec r_i \right)^2 \right\}\ .
\end{eqnarray}

\noindent By using $b$ as our unit of length and rewriting
Eq. (\ref{Hmc_viend}) in terms of dimensionless quantities
$\vec{y}_{(i)}=\vec{r}_i/b$ and $\vec{\nabla}_{(i)}=b\vec{\nabla}_i$,
one obtains

\begin{eqnarray}
H_{\mbox{\small c.m.\vphantom{intr}}}
= \textstyle
\frac{\hbar\omega}{2A}
\displaystyle\sum_{i,j=1}^A \left\{-\vec\nabla_{(i)}\vec\nabla_{(j)} +
\vec y_{(i)}\vec y_{(j)} \right\}\ .
\end{eqnarray}

\noindent In order to separate mixed terms one can use the identity valid
for an arbitrary entries $a_{ij}$ symmetrical with respect to the index
permutation $a_{ij}=a_{ji}$

\begin{eqnarray}
\sum_{i,j=1}^Aa_{ij}
= {\textstyle\frac{1}{A-1}}
\sum_{\stackrel{\scriptstyle i,j=1}{\scriptstyle i<j}}^A
[a_{ii} + 2(A-1)a_{ij} + a_{ij}]\ .
\end{eqnarray}

\noindent This leads to representation of
$H_{\mbox{\small c.m.\vphantom{intr}}}$ in terms of single-particle and
two-particle opera\-tors

\begin{eqnarray}
&&H_{\mbox{\small c.m.\vphantom{intr}}}
= {\textstyle\frac{\hbar\omega}{2}\frac{2}{A(A-1)}}
\sum\limits_{\stackrel{\scriptstyle i,j=1}{\scriptstyle i<j}}^A
\left\{ {\textstyle A{\frac{1}{2}}}
\left[-\vec\nabla_{(i)}^2 + \vec y_{(i)}^{\: 2} \right]
+ {\textstyle A{\frac{1}{2}}}
\left[-\vec\nabla_{(j)}^2 + \vec y_{(j)}^{\: 2} \right] \right.
\nonumber\\
&&- \left. (A-1){\textstyle\frac{1}{2}}
\left[ -(\vec\nabla_{(i)}-\vec\nabla_{(j)})^2
+(\vec y_{(i)}^{\: 2} - \vec y_{(j)}^{\: 2}) \right] \right\}\ .
\end{eqnarray}

\noindent Now, the matrix element of the centre-of-mass Hamiltonian
between the oscillator shell model functions assumes the form

\begin{eqnarray}  \label{Hmc_mat_0}
&&\langle EK{\mit\Delta}J{\mit\Pi}T|
{{\frac{H_{\mbox{\small c.m.\vphantom{j}}}}{{\hbar\omega}}}}
|EK^{\prime}{\mit\Delta}^{\prime}J{\mit\Pi}T\rangle
\nonumber\\
&&\displaystyle =
 {\textstyle\frac{1}{2}} \sum\limits_{\stackrel{\scriptstyle\overline{%
\overline{(EK{\mit\Delta}J{\mit\Pi}T)}}}{\scriptstyle J^{\prime\prime}T^{%
\prime\prime}}} \sum\limits_{\stackrel{\scriptstyle (elj)_{A-1},(elj)_A}{%
\scriptstyle (elj)_{A-1}^{\prime},(elj)_A^{\prime}}} \langle \overline{%
\overline{(EK{\mit\Delta}J{\mit\Pi}T)}};
((elj)_{A-1},(elj)_A)J^{\prime\prime}T^{\prime\prime}|| EK{\mit\Delta}J{\mit%
\Pi}T\rangle
\nonumber\\
&&\times \langle \overline{\overline{(EK{\mit\Delta}J{\mit\Pi}T)}};
((elj)_{A-1}^{\prime},(elj)_A^{\prime})J^{\prime\prime}T^{\prime\prime}||
EK^{\prime}{\mit\Delta}^{\prime}J{\mit\Pi}T\rangle
\nonumber\\
&&\times \langle
((elj)_{A-1},(elj)_A)J^{\prime\prime}T^{\prime\prime}|\; \left\{ A
{\textstyle\frac{1}{2}}
\left[-\vec\nabla_{(A-1)}^2 + \vec y_{(A-1)}^{\: 2} \right] + A
{\textstyle\frac{1}{2}}
\left[-\vec\nabla_{(A)}^2 + \vec y_{(A)}^{\: 2} \right]\right.
\nonumber\\
&&\left. -(A-1){\textstyle\frac{1}{2}} \left[ - (\vec\nabla_{(A-1)}-
\vec\nabla_{(A)})^2 + (\vec y_{(A-1)} - \vec y_{(A)})^2 \right] \right\}
|((elj)_{A-1}^{\prime},(elj)_A^{\prime})J^{\prime\prime}T^{\prime\prime}
\rangle\ .
\hphantom{55}
\end{eqnarray}

\noindent Here the usual shell model fractional parentage expansion
is used to separate out two last particles in the list, and the
notation follows that introduced in Ref. \cite{LF96}.
The terms corresponding to the harmonic oscillator operators of
one-particle type may be easily obtained by means of the
eigenvalue equation

\abovedisplayskip=0pt
\begin{eqnarray}
{\textstyle{\frac{1}{2}}}
\left( -\vec\nabla^2 + \vec\zeta^{\: 2} \right)
\phi_{\tilde e}(\vec\zeta\:)
= \left(\tilde e+{\textstyle{\frac{3}{2}}}\right)
\phi_{\tilde e}(\vec\zeta\:)\ .
\end{eqnarray}

\noindent Here $\phi_{\tilde e}(\vec\zeta\:)$ is the single-particle
harmonic oscillator function in argument $\vec\zeta$ with oscillator quantum
number $\tilde e$.
So, since many-particle oscillator shell model functions can be expressed in
terms of single-particle harmonic oscillator functions, the following equality
is rather straightforward

\begin{eqnarray}  \label{viend_operat}
&&\sum\limits_{\stackrel{\scriptstyle\overline{\overline{(EK{\mit\Delta}J{\mit\Pi}T)}}}
{\scriptstyle J^{\prime\prime}T^{\prime\prime}}}
\sum\limits_{\stackrel{\scriptstyle (elj)_{A-1},(elj)_A}
{\scriptstyle (elj)_{A-1}^{\prime},(elj)_A^{\prime}}}
\langle \overline{\overline{(EK{\mit\Delta}J{\mit\Pi}T)}};
((elj)_{A-1},(elj)_A)J^{\prime\prime}T^{\prime\prime}
|| EK{\mit\Delta}J{\mit\Pi}T\rangle
\nonumber\\
&&\times
\langle \overline{\overline{(EK{\mit\Delta}J{\mit\Pi}T)}};
((elj)_{A-1}^{\prime},(elj)_A^{\prime})J^{\prime\prime}T^{\prime\prime}
||EK^{\prime}{\mit\Delta}^{\prime}J{\mit\Pi}T\rangle
\nonumber\\
&&\vphantom{\Biggl(} \times
\langle ((elj)_{A-1},(elj)_A)J^{\prime\prime}T^{\prime\prime}
|\; \left\{A{\textstyle\frac{1}{2}}
\left[-\vec\nabla_{(A-1)}^2 + \vec y_{(A-1)}^{\: 2} \right]
+ A{\textstyle\frac{1}{2}}
\left[-\vec\nabla_{(A)}^2 + \vec y_{(A)}^{\: 2} \right]\right\}
\nonumber\\
&&\vphantom{\Biggl(}
|((elj)_{A-1}^{\prime},(elj)_A^{\prime})J^{\prime\prime}T^{\prime\prime}\rangle
\nonumber\\
&&=
\sum\limits_{\stackrel{\scriptstyle\overline{\overline{(EK{\mit\Delta}J{\mit\Pi}T)}}}
{\scriptstyle J^{\prime\prime}T^{\prime\prime}}}
\sum\limits_{(elj)_{A-1},(elj)_A}
\left[ A(e_{\scriptscriptstyle A-1}+{\textstyle\frac{3}{2}})
+ A(e_{\scriptscriptstyle A}+{\textstyle\frac{3}{2}}) \right]
\nonumber\\
&&\vphantom{\Biggl(} \times
\langle \overline{\overline{(EK{\mit\Delta}J{\mit\Pi}T)}};
((elj)_{A-1},(elj)_A)J^{\prime\prime}T^{\prime\prime}
|| EK{\mit\Delta}J{\mit\Pi}T\rangle
\nonumber\\
&&\vphantom{\Biggl(} \times
\langle \overline{\overline{(EK{\mit\Delta}J{\mit\Pi}T)}};
((elj)_{A-1},(elj)_A)J^{\prime\prime}T^{\prime\prime}
||EK^{\prime}{\mit\Delta}^{\prime}J{\mit\Pi}T\rangle\ .
\end{eqnarray}

The expectation value of the two-particle operators from Eq. (\ref{Hmc_mat_0})
can be obtained by expressing them into the single-particle form.
This may be accomplished by means of Jacobi coordinates

\begin{eqnarray}
\cases{
\vec\xi_{-1}
= {\textstyle\frac{1}{\sqrt{2}}} (\vec y_{(A-1)}+\vec y_{(A)})
\vphantom{\Biggl(} \cr
\vec\xi_{2}
= {\textstyle\frac{1}{\sqrt{2}}} (\vec y_{(A-1)}-\vec y_{(A)})\ .
\vphantom{\Biggl(} }
\end{eqnarray}

\noindent With their use one can rewrite the operator as

\begin{eqnarray}
{\textstyle{\frac{1}{2}}}
\left[ - (\vec\nabla_{(A-1)}-\vec\nabla_{(A)})^2
+ (\vec y_{(A-1)} - \vec y_{(A)})^2 \right]
= {\textstyle 2\; {\frac{1}{2}}}
\left( -\vec\nabla_{(2)}^2 + \vec\xi_{(2)}^{\: 2} \right)\ .
\end{eqnarray}

The same transformation should be accomplished for the wave
functions too.
The antisymmetric two-particle oscillator shell model functions may be expanded
in terms of vector coupled products of the $\psi_{elj\pi tm_jm_t}(\xi_2)$
functions depending on the intrinsic Jacobi variable $\xi_2$
with the $\psi_{(el)_{-1}}(\vec \xi_{-1})$ functions depending on the
Jacobi coordinate with the nonpositive index $\vec \xi_{-1}$ \cite{LF96}

\begin{eqnarray}
&&|((elj)_{A-1},(elj)_A)J^{\prime\prime}T^{\prime\prime}\rangle
=\displaystyle \sum\limits_{(el)_{-1},elj\pi t}
|((el)_{-1},elj\pi t)J^{\prime\prime}T^{\prime\prime}\rangle
\nonumber\\
&&\vphantom{\Biggl(} \times
\langle ((elj)_{A-1},(elj)_A)J^{\prime\prime}T^{\prime\prime}
||((el)_{-1},elj\pi t)J^{\prime\prime}T^{\prime\prime}\rangle\ .
\end{eqnarray}

\noindent Here $elj\pi tm_jm_t$ characterize the separated two-particle
subsystem, and Jacobi coordinate with nonpositive index is proportional to
the centre-of-mass coordinate of this separated two-particle subsystem.
Now, the functions $\psi _{elj\pi tm_jm_t}(\xi _2)$ are
eigenfunctions of the operator ${\textstyle {\frac 12}}\left( -\vec{\nabla}%
_{(2)}^2+\vec{\xi}_{(2)}^{:2}\right) $ as well as orthogonal and normalized,
thus for the two-particle term of Eq. (\ref{Hmc_mat_0}) one has

\begin{eqnarray}  \label{dvid_per_Jakob}
&&\langle ((elj)_{A-1},(elj)_A)J^{\prime\prime}T^{\prime\prime}|
\; {\textstyle (A-1){\frac{1}{2}}}
\left[ - (\vec\nabla_{(A-1)}-\vec\nabla_{(A)})^2
+ (\vec y_{(A-1)} - \vec y_{(A)})^2 \right]
\nonumber\\
&&|((elj)_{A-1}^{\prime},(elj)_A^{\prime})J^{\prime\prime}T^{\prime\prime}\rangle
\nonumber\\
&&\vphantom{\Biggl(} = \displaystyle
\sum\limits_{\stackrel{\scriptstyle{(el)_{-1},elj\pi t}}
{\scriptstyle{(el)_{-1}^{\prime},(elj\pi t)^{\prime}}} }
\langle((elj)_{A-1},(elj)_A)J^{\prime\prime}T^{\prime\prime}
|((el)_{-1},elj\pi t)J^{\prime\prime}T^{\prime\prime}\rangle
\nonumber\\
&&\vphantom{\Biggl(} \times
\langle((elj)_{A-1}^{\prime},(elj)_A^{\prime})J^{\prime\prime}T^{\prime\prime}
|((el)_{-1}^{\prime},(elj\pi t)^{\prime})J^{\prime\prime}T^{\prime\prime}\rangle
\nonumber\\
&&\vphantom{\Biggl(} \times
{\textstyle 2(A-1)} \langle((el)_{-1},elj\pi t)J^{\prime\prime}T^{\prime\prime}
| {\textstyle {\frac{1}{2}}}
\left( -\vec\nabla_{(2)}^2 + \vec\xi_{(2)}^{\: 2} \right)
|((el)_{-1}^{\prime},(elj\pi t)^{\prime})J^{\prime\prime}T^{\prime\prime}\rangle
\nonumber\\
&&\vphantom{\Biggl(} = \displaystyle
\sum\limits_{(el)_{-1},elj\pi t}
{\textstyle 2(A-1)}
\left( e+{\textstyle{\frac{3}{2}}}\right)
\langle((elj)_{A-1},(elj)_A)J^{\prime\prime}T^{\prime\prime}
|((el)_{-1},elj\pi t)J^{\prime\prime}T^{\prime\prime}\rangle
\nonumber\\
&&\vphantom{\Biggl(} \times
\langle((elj)_{A-1}^{\prime},(elj)_A^{\prime})J^{\prime\prime}T^{\prime\prime}
|((el)_{-1},elj\pi t)J^{\prime\prime}T^{\prime\prime}\rangle\ .
\end{eqnarray}

\noindent Here $e$ is the number of oscillator quanta corresponding to the
intrinsic motion of the separated two-particle subsystem.
It is convenient to extract the coefficient\\
${\textstyle 2(A-1)}\left( e+{\textstyle{\frac{3}{2}}}\right)$ out of the
Eq. (\ref{dvid_per_Jakob}) and move this coefficient to the right-hand
side of Eq. (\ref{viend_operat}).
After that it obtains the form

\begin{eqnarray}  \label{prijungta}
&&\displaystyle
\sum\limits_{\stackrel{\scriptstyle\overline{\overline{(EK{\mit\Delta}J{\mit\Pi}T)}}}
{\scriptstyle J^{\prime\prime}T^{\prime\prime}}}
\sum\limits_{(elj)_{A-1},(elj)_A} \left[ A(e_{\scriptscriptstyle A-1}
+e_{\scriptscriptstyle A})+3 \right]
\nonumber\\
&&\vphantom{\Biggl(} \times
\langle \overline{\overline{(EK{\mit\Delta}J{\mit\Pi}T)}};
((elj)_{A-1},(elj)_A)J^{\prime\prime}T^{\prime\prime}
|| EK{\mit\Delta}J{\mit\Pi}T\rangle
\nonumber\\
&&\vphantom{\Biggl(} \times
\langle \overline{\overline{(EK{\mit\Delta}J{\mit\Pi}T)}};
((elj)_{A-1},(elj)_A)J^{\prime\prime}T^{\prime\prime}
|| EK^{\prime}{\mit\Delta}^{\prime}J{\mit\Pi}T\rangle\ .
\end{eqnarray}

Taking into account Eqs. (\ref{dvid_per_Jakob}) and (\ref{prijungta}) the
centre-of-mass Hamiltonian matrix element becomes

\begin{eqnarray}  \label{Hcm_priespask}
&&\langle EK{\mit\Delta}J{\mit\Pi}T |
\displaystyle{\frac{H_{\mbox{\small c.m.\vphantom{intr}}}}{{\hbar\omega}}}
| EK^{\prime}{\mit\Delta}^{\prime}J{\mit\Pi}T\rangle
\nonumber\\
&&\vphantom{\Biggl(} = \displaystyle
\sum\limits_{\stackrel{\scriptstyle\overline{\overline{(EK{\mit\Delta}J{\mit\Pi}T)}}}
{\scriptstyle J^{\prime\prime}T^{\prime\prime}} }
\sum\limits_{\stackrel{\scriptstyle (elj)_{A-1},(elj)_A}{\scriptstyle (elj)_{A-1}^{\prime},(elj)_A^{\prime}} }
\langle\overline{\overline{(EK{\mit\Delta}J{\mit\Pi}T)}};
((elj)_{A-1},(elj)_A)J^{\prime\prime}T^{\prime\prime}
|| EK{\mit\Delta}J{\mit\Pi}T\rangle
\nonumber\\
&&\vphantom{\Biggl(} \times
\langle\overline{\overline{(EK{\mit\Delta}J{\mit\Pi}T)}};
((elj)_{A-1}^{\prime},(elj)_A^{\prime})J^{\prime\prime}T^{\prime\prime}
||EK^{\prime}{\mit\Delta}^{\prime}J{\mit\Pi}T\rangle
\nonumber\\
&&\vphantom{\Biggl(} \times
\Bigl\{ [{\textstyle\frac{A}{2}}(e_{\scriptscriptstyle A-1}+e_{\scriptscriptstyle A})
+{\textstyle\frac{3}{2}})]
\delta_{(elj)_{A-1},(elj)_{A-1}^{\prime}}\delta_{(elj)_A,(elj)_A^{\prime}}
\nonumber\\
&&\vphantom{\Biggl(}
- \, (A-1) \displaystyle\sum\limits_{elj\pi t,(el)_{-1}} e\;\;
\langle((elj)_{A-1},(elj)_A)J^{\prime\prime}T^{\prime\prime}
|((el)_{-1},elj\pi t)J^{\prime\prime}T^{\prime\prime}\rangle
\nonumber\\
&&\vphantom{\Biggl(} \times
\langle((elj)_{A-1}^{\prime},(elj)_A^{\prime})J^{\prime\prime}T^{\prime\prime}
|((el)_{-1},elj\pi t)J^{\prime\prime}T^{\prime\prime}\rangle \Bigr\}
\ .
\end{eqnarray}

It is possible to express the sum over the single-particle energy in terms
of the total energy $E$.
For this, let us express the total Hamiltonian,
defined as the sum of single-particle operators, in terms of the sum of
two-particle operators

\begin{eqnarray}
{\displaystyle {\frac{H}{{\hbar\omega}}} }
= {\displaystyle\sum_{i = 1}^A} {\textstyle{\frac{1}{2}}}
\left\{ - \vec\nabla_{(i)}^2 + \vec y{\hphantom{\,}}_{(i)}^2 \right\}
= {\displaystyle\sum_{i = 1}^A} h_{(i)}
= {\textstyle{\frac{1}{{A-1}}}} \displaystyle
\sum_{\stackrel{\scriptstyle i,j=1}{\scriptstyle i<j}}^A
\left( h_{(i)} + h_{(j)} \right)\ .
\end{eqnarray}

\noindent Now, the total Hamiltonian matrix element between the oscillator shell
model functions assumes the form

\begin{eqnarray}  \label{H_per_dvid}
&&\langle EK{\mit\Delta}J{\mit\Pi}T
| { \displaystyle {\frac{H}{{\hbar\omega}}} }
| EK^{\prime}{\mit\Delta}^{\prime}J{\mit\Pi}T\rangle
\nonumber\\
&& = {\textstyle {\frac{1}{{A-1} }} } {\textstyle {A \choose 2}}
\langle EK{\mit\Delta}J{\mit\Pi}T | ( h_{(A-1)} + h_{(A)} )
| EK^{\prime}{\mit\Delta}^{\prime}J{\mit\Pi}T\rangle
\nonumber\\
&&\vphantom{\Biggl(}
= {\displaystyle
\sum\limits_{\stackrel{\scriptstyle\overline{\overline{(EK{\mit\Delta}J{\mit\Pi}T)}}}
{\scriptstyle J^{\prime\prime}T^{\prime\prime}} }
\sum\limits_{(elj)_{A-1},(elj)_A} }
{\textstyle\frac{A}{2}}(e_{\scriptscriptstyle A-1}+{\textstyle\frac{3}{2}}
+ e_{\scriptscriptstyle A}+{\textstyle\frac{3}{2}})
%\hphantom{123..}
\nonumber\\
&&\vphantom{\Biggl(} \times
\langle\overline{\overline{(EK{\mit\Delta}J{\mit\Pi}T)}};
((elj)_{A-1},(elj)_A)J^{\prime\prime}T^{\prime\prime}
|| EK{\mit\Delta}J{\mit\Pi}T\rangle
\nonumber\\
&&\vphantom{\Biggl(} \times
\langle\overline{\overline{(EK{\mit\Delta}J{\mit\Pi}T)}};
((elj)_{A-1},(elj)_A)J^{\prime\prime}T^{\prime\prime}
|| EK^{\prime}{\mit\Delta}^{\prime}J{\mit\Pi}T\rangle\ .
\end{eqnarray}

The oscillator shell model functions are the eigenfunctions of the
total Hamiltonian, so the corresponding matrix element is by
definition

\begin{eqnarray}  \label{Hnuos_E}
\langle EK{\mit\Delta}J{\mit\Pi}T
| {\ \displaystyle {\frac{H}{{\hbar\omega}}} }
| EK^{\prime}{\mit\Delta}^{\prime}J{\mit\Pi}T\rangle
= \left(E+{\textstyle {\frac{3}{2}}}A\right)
\delta_{K{\mit\Delta},K^{\prime}{\mit\Delta}^{\prime}}\ .
\end{eqnarray}

Comparing Eqs. (\ref{H_per_dvid}) and (\ref{Hnuos_E}) one obtains

\begin{eqnarray}  \label{E_per_vnd}
&& E\cdot\delta_{K{\mit\Delta},K^{\prime}{\mit\Delta}^{\prime}}
= {\displaystyle
\sum\limits_{\stackrel{\scriptstyle\overline{\overline{(EK{\mit\Delta}J{\mit\Pi}T)}}}
{\scriptstyle J^{\prime\prime}T^{\prime\prime}} }
\sum\limits_{(elj)_{A-1},(elj)_A} } {\textstyle\frac{A}{2}}
(e_{\scriptscriptstyle A-1}+e_{\scriptscriptstyle A})
\nonumber\\
&&\vphantom{\Biggl(} \times
\langle\overline{\overline{(EK{\mit\Delta}J{\mit\Pi}T)}};
((elj)_{A-1},(elj)_A)J^{\prime\prime}T^{\prime\prime}
|| EK{\mit\Delta}J{\mit\Pi}T\rangle
\nonumber\\
&&\vphantom{\Biggl(} \times
\langle\overline{\overline{(EK{\mit\Delta}J{\mit\Pi}T)}};
((elj)_{A-1},(elj)_A)J^{\prime\prime}T^{\prime\prime}
|| EK^{\prime}{\mit\Delta}^{\prime}J{\mit\Pi}T\rangle\ .
\end{eqnarray}

Inserting in the equality (\ref{Hcm_priespask}) the expression (\ref
{E_per_vnd}) and introducing the coefficients of expansion of the oscillator
shell model functions in terms of the ones with singled out dependence on the
intrinsic coordinates of two last particles (SCFP's), we get the final
expression for the centre-of-mass Hamiltonian matrix element

\begin{eqnarray}
&&\langle EK{\mit\Delta}J{\mit\Pi}T |
{\displaystyle{\frac{H_{\mbox{\small c.m.\vphantom{intr}}}}{{\hbar\omega}}}}
| EK^{\prime}{\mit\Delta}^{\prime}J{\mit\Pi}T\rangle
= \left(E + {\textstyle\frac{3}{2}}\right) \,
\delta_{K{\mit\Delta},K^{\prime}{\mit\Delta}^{\prime}} - \, (A-1)
\displaystyle\sum\limits_{elj\pi t} e
\nonumber\\
&&\vphantom{\Biggl(}
\times \displaystyle \sum\limits_{(el)_{-1},J^{\prime\prime}T^{\prime\prime}}
\sum\limits_{\overline{\overline{(EK{\mit\Delta}J{\mit\Pi}T)}}}
\langle\overline{\overline{(EK{\mit\Delta}J{\mit\Pi}T)}};
((el)_{-1},elj\pi t)J^{\prime\prime}T^{\prime\prime}
|| EK{\mit\Delta}J{\mit\Pi}T\rangle
\nonumber\\
&&\vphantom{\Biggl(}
\times \langle\overline{\overline{(EK{\mit\Delta}J{\mit\Pi}T)}};
((el)_{-1},elj\pi t)J^{\prime\prime}T^{\prime\prime}
|| EK^{\prime}{\mit\Delta}^{\prime}J{\mit\Pi}T\rangle\ .
\end{eqnarray}

\section{Eigenvector problem of the real symmetric matrix}

For the centre-of-mass Hamiltonian matrix obtained, the general
eigenvalue problem should be solved.
Usually, this is carried out by some numerical methods.
However, numerical methods are inefficient for highly degenerate subspaces,
what is the case.
The problem at hand is really less complicated, since the eigenvalues of the
matrix under consideration are known.
Moreover, this problem can be generalized, since eigenvalues of the quadratic
Casimir operators for almost all physical interesting groups are well defined.
So, the problem can be formulated as to obtain the eigenvectors of the real
symmetric matrix ${\bf M}$, when its eigenvalues are known.
We present here a new method which allows one to find the analytical expressions
of the orthonormalyzed system of eigenvectors, corresponding to the chosen
eigenvalue of the matrix.
Since the symmetrical properties of functions are usually formulated in terms
of eigenvalues of the quadratic Casimir operators, this method makes it
possible to obtain the set of functions with necessary symmetrical
characteristics.
Let us outline shortly the essence of the method.

It is well-known that the real symmetric matrix ${\bf M}$ is expressible as the
spectral decomposition of idempotent matrices \cite{Lankaster69}

%\addtocounter{numeris}{1}
\begin{eqnarray}
{\bf M}=\sum_{\alpha=1}^s \lambda_\alpha {\bf P}_\alpha\ .
\end{eqnarray}

\noindent Here $({\bf P}_\alpha)_{n\times n}$ is the idempotent matrix of
the rank $r_\alpha$, and $\lambda_\alpha$ denote the distinct eigenvalues
of the matrix ${\bf M}$.
Idempotent matrices are known to obey the condition

%\addtocounter{numeris}{1}
\begin{eqnarray}
{\bf P}_\alpha\cdot{\bf P}_\beta = \delta_{\alpha\beta}{\bf P}_\alpha\ .
\end{eqnarray}

Using the Cayley-Hamilton theorem \cite{Lankaster69}, the matrix
${\bf P}_\alpha$ may be expressed in terms of the original matrix ${\bf M}$ and
its eigenvalues

\begin{eqnarray}
{\bf P}_\alpha
= \prod_{\stackrel{\scriptstyle \beta=1}{\scriptstyle \beta\ne\alpha}}^s
(\lambda_\beta - {\bf M})
/ \prod_{\stackrel{\scriptstyle \beta=1}{\scriptstyle \beta\ne\alpha}}^s
(\lambda_\beta - \lambda_\alpha) \ .
\end{eqnarray}

It is known, that the spectral decomposition of an idempotent matrix is
\cite{Lankaster69}

%\addtocounter{numeris}{1}
\begin{eqnarray}
{\bf P}_\alpha =\tilde{{\bf F}}_\alpha\cdot\tilde{{\bf F}}_\alpha^{+}\ ,
\end{eqnarray}

\noindent where the matrix $\tilde{{\bf F}}_\alpha$ is composed of
eigenvectors of the matrix {\bf P}$_\alpha$, corresponding to its
eigenvalues equal to one.
In the case of real symmetric matrices, $\tilde{{\bf F}}_\alpha^{+}$ is the
transposed matrix.
The eigenvectors satisfy the usual normalization condition
$\tilde{{\bf F}}_\alpha^+\cdot\tilde{{\bf F}}_\alpha = 1$.

Taking into account the above spectral decomposition of the matrix ${\bf P}%
_\alpha$ it is straightforward to show that

%\addtocounter{numeris}{1}
\begin{eqnarray}
{\bf M}\cdot\tilde{{\bf F}}_\alpha = \tilde{{\bf F}}_\alpha\cdot\lambda_%
\alpha \ .
\end{eqnarray}

\noindent It means that columns of the matrix $\tilde{{\bf F}}_\alpha$ are
eigenvectors of the matrix ${\bf M}$ corresponding to the eigenvalue
$\lambda_\alpha$.

The problem of constructing the eigenvectors for an idempotent matrix was
once resolved in the case of coefficients of fractional parentage
for antisymmetric wave functions \cite{LF95}.
The method is based on the observation that the spectral decomposition of the
matrix ${\bf P}$ is defined only within an orthogonal transformation.
So, the matrix {\bf F} defined by

%\addtocounter{numeris}{1}
\begin{eqnarray}
{\bf F} = \tilde{{\bf F}}\cdot{\bf G}\ ,
\end{eqnarray}

\noindent also gives the spectral decomposition of the matrix ${\bf P}$.
The orthogonal matrix {\bf G} has $r(r-1)/2$ independent parameters, they may
be chosen in a way which allows one to fix the corresponding number of the matrix
{\bf F} elements.
The best choice is

%\addtocounter{numeris}{1}
\begin{eqnarray}
(F_\alpha)_{ij} = 0\; \quad\hbox{\rm if} \quad 1\leq i < j \leq r_\alpha\ .
\end{eqnarray}

\noindent In such a case the expressions for eigenvectors {\bf F}$_\alpha$
of the real symmetric matrix ${\bf M}$ corresponding to the eigenvalue $%
\lambda_\alpha$ take the form

\begin{eqnarray}  \label{eigenvectors}
  \left\{
    \begin{array}{lll}
\displaystyle
(F_\alpha)^{\, 2}_{11} & = & \displaystyle (P_\alpha)_{11} \ ,
\\
\displaystyle
(F_\alpha)_{j1} & = & \displaystyle {\frac{(P_\alpha)_{1j}}{(F_\alpha)_{11}}} \ ,
\\
\displaystyle
(F_\alpha)^{\, 2}_{ii} &
= & \displaystyle (P_\alpha)_{ii}-\sum_{k=1}^{i-1}(F_\alpha)_{ik}^{\, 2} \ ,
\\
\displaystyle
(F_\alpha)_{ji} & = & \displaystyle {\frac{1}{(F_\alpha)_{ii}}} \left\{
(P_\alpha)_{ij}-\sum_{k=1}^{i-1}(F_\alpha)_{ik}(F_\alpha)_{jk} \right\}\ ,
    \end{array}
  \right.
\end{eqnarray}

\noindent for every value of $i = 2, 3,\dots , r$ and the corresponding set of
$j = i+1, i+2,\ldots , n$.
Here $n$ is the dimension and $r$ is the rank of the matrix ${\bf P}_{\alpha}$.
The $r$ columns of the matrix {\bf F}$_\alpha$ may be obtained from the set of
any $r$ linearly independent rows of the matrix ${\bf P}_{\alpha}$.
Consider the first nonzero row of the matrix ${\bf P}_{\alpha}$.
Let it be the $k$-th.
Then the calculation begins with this $k$-th row of the matrix
${\bf P}_{\alpha}$: $(F_\alpha)^{\, 2}_{11} = (P_\alpha)_{kk}$ and
$(F_\alpha)_{j1} = {\frac{(P_\alpha)_{kj}}{(F_\alpha)_{11}}}$.
Because the matrix ${\bf P}_{\alpha}$ is symmetric, every next $i$-th row
calculation starts from the diagonal element $(F_\alpha)_{ii}$.
This $i$-th row of the matrix ${\bf P}_{\alpha}$ will be linearly independent
of rows just calculated if the obtained diagonal element
$(F_\alpha)_{ii}\ne 0$.
In this case the corresponding column of the matrix {\bf F}$_\alpha$ could be
obtained.
If this condition is not fulfilled, the next row of the matrix
${\bf P}_{\alpha}$ should be tested.
Since the matrix ${\bf P}_{\alpha}$ has rank $r$, it will surely contain $r$
linearly  independent rows.
Positive values of $F_{ii}$ are convenient, because the overall sign of the
eigenvector is arbitrary.

The outlined method for constructing the eigenvectors with necessary
characteristics can be applied for diagonalization of the $H_{\mbox{\small
c.m.\vphantom{intr}}}$ matrix.
The obtained eigenvectors {\bf F} will be the sets of coefficients
$a_{K{\mit\Delta };00,{\mit\Gamma }}^{EJ{\mit\Pi }T}$ introduced in Ref.
\cite{LF96}.
Then the coefficients of expansion of the oscillator shell model functions in
terms of $A$-particle oscillator functions with singled out dependence on
intrinsic coordinates of two last particles and with eliminated spurious states
(CESO's), as well as the intrinsic density matrices can be calculated.

\section{Numerical benchmarks}

Let us use as a benchmark the formation of the matrix $\left\langle\;H_{%
\mbox{\small c.m.\vphantom{intr}}}\;\right\rangle$, its diagonalization, and
calculation of the CESO's and intrinsic density matrices for the $A=3$ case.

The eigenvalues of the $H_{\mbox{\small c.m.\vphantom{intr}}}$ matrix makes
the sequence ${\frac{3}{2}}, {\frac{5}{2}}\ldots E - E_{\min} + {\frac{3}{2}}
$, where $E$ is the total oscillator energy value of the state.

Let us start from the case $E = E_{\min}$. Then, for three particles there
is only one configuration $K_1$, and the total angular momentum as well as
the isospin can take only one value

$K_1\equiv\left(00{\frac{1}{2}}\right)^3$: \hphantom{12} $J{\mit\Pi}T = {%
\frac{1}{2}}^+{\frac{1}{2}}\ .$

The state (1) in the configuration space will be denoted by the sequence of
sing\-le shells $(elj)^n$ accompanied with $JT$ values in correspondence with
the momentum coupling scheme.

$(1)\quad\Bigl(\left(00{\frac{1}{2}}\right)^3\Bigr){\frac{1}{2}}{\frac{1}{2}}
\ .$

In the case when $E = E_{\min}$, the $J$ and $T$ could take only the
displayed values.
For brevity, let us deal below with the same
$J{\mit\Pi}T = {\frac{1}{2}}^+{\frac{1}{2}}$ values also for higher total
oscillator energies.

In the case of three particles and $E = E_{\min} = 0$ the matrix of the
operator $H_{\mbox{\small c.m.\vphantom{intr}}}$ is the single number

\begin{eqnarray*}
\left\langle\;
{\displaystyle{\frac{H_{\mbox{\small c.m.\vphantom{intr}}} }{{\hbar\omega}}}}
\;\right\rangle_{\scriptstyle A=3,E=0,J{\mit\Pi}T={\frac{1}{2}}^+{\frac{1}{2}}}
= {\textstyle\frac{3}{2}}\ .
\end{eqnarray*}

\noindent This is why the oscillator shell model states characterized by
$E_{\min}$ contain the centre of mass in its ground state and all
$a_{K{\mit\Delta};00,{\mit\Gamma}}^{EJ{\mit\Pi}T} = 1$.

Now the centre of mass is not excited, hence the CESO's coincide with
the \\ coefficients of expansion of the oscillator shell model function in
terms of the ones with singled out dependence on intrinsic coordinates of two
last particles.
They are presented in Table~1.
Here the more usual spectroscopic notation $\raisebox{6pt}{\scriptsize
2s+1}l_{\scriptstyle j}$ is used to denote the state of a subsystem
containing two separated particles.

The intrinsic density matrices, in the case when $A = 3,\; E=0,\; JT={\frac{%
1}{2}}{\frac{1}{2}}$ are

\begin{eqnarray*}
\begin{array}{c}
{\bf W}(\,\raisebox{6pt}{\scriptsize 1}S_0) = \left(
\begin{array}{r}
\frac{1}{2}
\end{array}
\right)
\end{array}
\ ,
\begin{array}{c}
{\bf W}(\,\raisebox{6pt}{\scriptsize 3}S_1) = \left(
\begin{array}{r}
\frac{1}{2}
\end{array}
\right)
\end{array}
\ .
\end{eqnarray*}

In the case of $E = E_{\min} + 1 = 1$, there are two configurations and a
number of total angular momentum and isospin values appear:

\begin{tabbing}
$K_1\equiv\left(00{1\over2}\right)^2\left(11{1\over2}\right)$:
\hphantom{12}
\= $J{\mit\Pi}T = {1\over2}^-{1\over2},\;  {1\over2}^-{3\over2},\;
{3\over2}^-{1\over2}$, \\
$K_2\equiv\left(00{1\over2}\right)^2\left(11{3\over2}\right)$: \>
\= $J{\mit\Pi}T = {1\over2}^-{1\over2},\;  {3\over2}^-{1\over2},\;
{3\over2}^-{3\over2},\;  {5\over2}^-{1\over2}$. \\
\end{tabbing}

Now, the $J{\mit\Pi}T = {\frac{1}{2}}^+{\frac{1}{2}}$ values may be obtained
for the three states in the configuration space:

$(1)\quad\Bigl(\left(00{\frac{1}{2}}\right)^201,\left(11{\frac{1}{2}}\right)
{\frac{1}{2}}{\frac{1}{2}}\Bigr){\frac{1}{2}}{\frac{1}{2}}$,

$(2)\quad\Bigl(\left(00{\frac{1}{2}}\right)^210,\left(11{\frac{1}{2}}\right)
{\frac{1}{2}}{\frac{1}{2}}\Bigr){\frac{1}{2}}{\frac{1}{2}}$,

$(3)\quad\Bigl(\left(00{\frac{1}{2}}\right)^210,\left(11{\frac{3}{2}}\right)
{\frac{3}{2}}{\frac{1}{2}}\Bigr){\frac{1}{2}}{\frac{1}{2}}$. \vspace{0.3 cm}

In order to calculate the $\left\langle \;H_{%
\mbox{\small
c.m.\vphantom{intr}}}\;\right\rangle $ matrix, the coefficients of expansion
of the oscillator shell model functions in terms of the ones with singled out
dependence on intrinsic coordinates of two last particles for $E=E_{\min }+1$
should be known.
They are presented in Table~2. Here only the values of the indices
different from the preceding row are shown.

Since there are three states in the configuration space, the dimension of the
matrix of the operator $H_{\mbox{\small c.m.\vphantom{intr}}}$ will be three:

\begin{eqnarray*}
\left\langle\; {\displaystyle{\frac{H_{\mbox{\small c.m.\vphantom{intr}}} }{{%
\hbar\omega}}}} \;\right\rangle_{\scriptstyle A=3,E=1,J{\mit\Pi}T={\frac{1}{2%
}}^-{\frac{1}{2}}} = \left( \matrix{ \hphantom{12}2&\frac{1}{6}
&\frac{2}{3\sqrt{2}}& \cr &\frac{14}{9}&\frac{2}{9\sqrt{2}}& \cr &
&\frac{35}{18} & \cr } \right)\ ,
\end{eqnarray*}

\noindent where the rows and columns of the matrix $\left\langle \;H_{%
\mbox{\small
c.m.\vphantom{intr}}}\;\right\rangle $ are labeled by the states in the
configuration space that are taken in the sequence (1), (2), (3). Due to the
symmetric nature of this matrix here we show only its upper triangle. The
rank of the idempotent matrices of this matrix is two, so there will be
only two normalized eigenvectors. After diagonalization of the matrix $%
\left\langle \;H_{\mbox{\small c.m.\vphantom{intr}}}\;\right\rangle $
according to the outlined above method and using Eq. (\ref{eigenvectors}),
one may obtain

\begin{eqnarray*}
\left\langle\; a_{K{\mit\Delta};00,{\mit\Gamma}}\;\right\rangle ^{%
\scriptstyle A=3,E=1,J{\mit\Pi}T={\frac{1}{2}}^-{\frac{1}{2}}} = \left( %
\matrix{ \frac{1}{\sqrt{2}}& 0& \cr
-\frac{1}{3\sqrt{2}}&\frac{4}{3\sqrt{2}}& \cr
-\frac{2}{3}&-\frac{1}{3}& \cr } \right)\ ,
\end{eqnarray*}

\noindent where rows of the matrix $\left\langle\; a_{K{\mit\Delta};00,{\mit%
\Gamma}}\;\right\rangle$ are labeled by the states in the configuration
space that are taken in the same sequence (1), (2), (3) as for $%
\left\langle\;H_{\mbox{\small c.m.\vphantom{intr}}}\;\right\rangle$, columns
are labeled by the additional integer quantum number ${\mit\Gamma}$ which
can take two values: 1 and 2, in correspondence with the rank of the matrix $%
\left\langle\;H_{\mbox{\small c.m.\vphantom{intr}}}\;\right\rangle$ or, in
another words, with the number of its linearly independent eigenvectors. It
should be stressed, that the empty place instead of the first element of the
second column means that its value is equal to zero, in full accordance with
prescriptions of the method.

The results of CESO's calculations are presented in Table 3.
It should be stressed, that CESO's are labeled with the set of indices:
$\overline{\overline{(EK{\mit\Delta }J{\mit\Pi }T)}}((el)_{-1},elj\pi t)J^
{\prime \prime}T^{\prime \prime }$ and the additional integer quantum number
${\mit\Gamma}$.
There are only different sets of them, but they can belong to all states
in the configuration space, regardless of the origin of the ground
configuration.
For example, if seven different sets of CESO's indices are taken
from the state (1), then only one can be taken from the state (2), since the
remaining six have the same sets of indices as in the state (1). At last, a
set of indices will be taken from the state (3), since the remaining four may
be met in states (1) and (3). Table 3 shows only the values of the indices
different from the preceding row, whereas CESO's which are absent here have
zero values.

Finally, by means of obtained CESO's the intrinsic density matrices can be
calculated.

The intrinsic density matrices in the case of $A = 3,\; E=1,\; JT={\frac{1}{2}%
}{\frac{1}{2}}$:

\begin{eqnarray*}
\begin{array}{c}
{\bf W}(\,\raisebox{6pt}{\scriptsize 1}S_0) = \left(
\begin{array}{r}
\frac{1}{4}
\end{array}
\right)
\end{array}
\ ,
\begin{array}{c}
{\bf W}(\,\raisebox{6pt}{\scriptsize 3}S_1) = \left(
\begin{array}{rr}
\frac{1}{4}\hphantom{1} & 0 \\
& \hphantom{1}\frac{1}{2}
\end{array}
\right)
\end{array}
\ ,
\end{eqnarray*}

\begin{eqnarray*}
\begin{array}{c}
{\bf W}(\,\raisebox{6pt}{\scriptsize 3}P_0) = \left(
\begin{array}{rr}
\frac{1}{12} & -\frac{1}{6} \\
& \frac{1}{3}
\end{array}
\right)
\end{array}
\ ,
\begin{array}{c}
{\bf W}(\,\raisebox{6pt}{\scriptsize 1}P_1) = \left(
\begin{array}{r}
\frac{1}{4}
\end{array}
\right)
\end{array}
\ ,
\begin{array}{c}
{\bf W}(\,\raisebox{6pt}{\scriptsize 3}P_1) = \left(
\begin{array}{rr}
\frac{1}{6}\hphantom{1} & \frac{1}{6} \\
& \hphantom{1}\frac{1}{6}
\end{array}
\right)
\end{array}
\ .
\end{eqnarray*}

\noindent The rows and columns of the intrinsic density matrices are labeled
by the additional integer quantum number ${\mit\Gamma }$, preserving
ascending order, so the dimensions of the matrices can vary from 1 to 2.
Only the upper triangle of the intrinsic density matrices is shown due to
its symmetry under index permutation.

In the case of $E = E_{\min} + 2 = 2$ there are six configurations that
produce a number of total angular momentum and isospin values:

\begin{tabbing}
$K_1\equiv\left(00{1\over2}\right)^2\left(20{1\over2}\right)$:
\hphantom{1234567}
\= $J{\mit\Pi}T = {1\over2}^+{1\over2},\;  {1\over2}^+{3\over2},\;
{3\over2}^+{1\over2}$, \\
$K_2\equiv\left(00{1\over2}\right)^2\left(22{3\over2}\right)$: \>
\= $J{\mit\Pi}T = {1\over2}^+{1\over2},\;  {3\over2}^+{1\over2},\;
{3\over2}^+{3\over2},\;  {5\over2}^+{1\over2}$, \\
$K_3\equiv\left(00{1\over2}\right)^2\left(22{5\over2}\right)$: \>
\= $J{\mit\Pi}T = {3\over2}^+{1\over2},\;  {5\over2}^+{1\over2},\;
{5\over2}^+{3\over2},\;  {7\over2}^+{1\over2}$, \\
$K_4\equiv\left(00{1\over2}\right)^2\left(11{1\over2}\right)^2$: \>
\= $J{\mit\Pi}T = {1\over2}^+{1\over2},\;  {1\over2}^+{3\over2},\;
{3\over2}^+{1\over2}$, \\
$K_5\equiv
\left(00{1\over2}\right)\left(11{1\over2}\right)\left(11{3\over2}\right)$:
\>
\= $J{\mit\Pi}T = {1\over2}^+{1\over2},\;  {1\over2}^+{3\over2},\;
{3\over2}^+{1\over2},\;  {3\over2}^+{3\over2},\;  {5\over2}^+{1\over2}
,\;  {5\over2}^+{3\over2}$, \\
$K_6\equiv\left(00{1\over2}\right)^2\left(11{1\over2}\right)^2$: \>
\= $J{\mit\Pi}T = {1\over2}^+{1\over2},\;  {1\over2}^+{3\over2},\;
{3\over2}^+{1\over2},\;  {3\over2}^+{3\over2},\;  {5\over2}^+{1\over2}
,\;  {5\over2}^+{3\over2},\;  {7\over2}^+{1\over2}$. \\
\end{tabbing}

Here and bellow the remarks, about notations mentioned above, will be valid,
so the data will be presented in more concise manner.

The $J{\mit\Pi}T = {\frac{1}{2}}^+{\frac{1}{2}}$ values could be obtained
for nine states in the configuration space:

$(1)\quad\Bigl(\left(00{\frac{1}{2}}\right)^201,\left(20{\frac{1}{2}}\right)
{\frac{1}{2}}{\frac{1}{2}}\Bigr){\frac{1}{2}}{\frac{1}{2}}$,

$(2)\quad\Bigl(\left(00{\frac{1}{2}}\right)^210,\left(20{\frac{1}{2}}\right)
{\frac{1}{2}}{\frac{1}{2}}\Bigr){\frac{1}{2}}{\frac{1}{2}}$,

$(3)\quad\Bigl(\left(00{\frac{1}{2}}\right)^210,\left(22{\frac{3}{2}}\right)
{\frac{3}{2}}{\frac{1}{2}}\Bigr){\frac{1}{2}}{\frac{1}{2}}$,

$(4)\quad\Bigl(\left(00{\frac{1}{2}}\right){\frac{1}{2}}{\frac{1}{2}}%
,\left(11{\frac{1}{2}}\right)^2 01\Bigr){\frac{1}{2}}{\frac{1}{2}}$,

$(5)\quad\Bigl(\left(00{\frac{1}{2}}\right){\frac{1}{2}}{\frac{1}{2}}%
,\left(11{\frac{1}{2}}\right)^2 10\Bigr){\frac{1}{2}}{\frac{1}{2}}$,

$(6)\quad\Bigl(\Bigl(\left(00{\frac{1}{2}}\right){\frac{1}{2}}{\frac{1}{2}},
\left(11{\frac{1}{2}}\right){\frac{1}{2}}{\frac{1}{2}}\Bigr)10, \left(11{%
\frac{3}{2}}\right){\frac{3}{2}}{\frac{1}{2}}\Bigr){\frac{1}{2}}{\frac{1}{2}}
$,

$(7)\quad\Bigl(\Bigl(\left(00{\frac{1}{2}}\right){\frac{1}{2}}{\frac{1}{2}},
\left(11{\frac{1}{2}}\right){\frac{1}{2}}{\frac{1}{2}}\Bigr)11, \left(11{%
\frac{3}{2}}\right){\frac{3}{2}}{\frac{1}{2}}\Bigr){\frac{1}{2}}{\frac{1}{2}}
$,

$(8)\quad\Bigl(\left(00{\frac{1}{2}}\right){\frac{1}{2}}{\frac{1}{2}},
\left(11{\frac{3}{2}}\right)^201\Bigr){\frac{1}{2}}{\frac{1}{2}}$,

$(9)\quad\Bigl(\left(00{\frac{1}{2}}\right){\frac{1}{2}}{\frac{1}{2}},
\left(11{\frac{3}{2}}\right)^210\Bigr){\frac{1}{2}}{\frac{1}{2}}$.

The number of the coefficients of expansion of the oscillator shell model
functions in terms of the
ones with singled out dependence on the intrinsic coordinates of two last
particles is 92, so they will not be presented here for the reasons of space.

Since there are nine states in the configuration space, the dimension of the
matrix of the operator $H_{\mbox{\small c.m.\vphantom{intr}}}$ will be nine %
\vspace{0.2 cm}

\begin{eqnarray*}
\left\langle {\displaystyle{\frac{H_{\mbox{\small c.m.\vphantom{intr}}} }{{%
\hbar\omega}}}} \right\rangle_{\scriptstyle A=3,E=2}^{\scriptstyle J{\mit\Pi}%
T={\frac{1}{2}}^+{\frac{1}{2}}} =
\left(
\begin{array}{rrrrrrrrr}
{\frac{13}{6}} & 0 & 0 & -\frac{1}{3\sqrt{6}} & -\frac{1}{3\sqrt{6}} & -%
\frac{2}{3\sqrt{6}} & \frac{2}{3\sqrt{2}} & -\frac{1}{3\sqrt{3}} & \frac{5}{3%
\sqrt{15}} \\
& \frac{13}{6} & 0 & \frac{1}{\sqrt{6}} & \frac{1}{9\sqrt{6}} & \frac{2}{9%
\sqrt{6}} & -\frac{2}{9\sqrt{2}} & \frac{1}{\sqrt{3}} & -\frac{5}{9\sqrt{15}}
\\
&  & \frac{13}{6} & 0 & \frac{20}{9\sqrt{30}} & -\frac{5}{9\sqrt{30}} &
\frac{5}{9\sqrt{10}} & 0 & -\frac{2}{9\sqrt{3}} \\
&  &  & \frac{7}{3} & \frac{1}{6} & 0 & -\frac{2}{3\sqrt{3}} & 0 & 0 \\
&  &  &  & \frac{17}{9} & \frac{2}{9} & 0 & 0 & 0 \\
&  &  &  &  & \frac{7}{3} & -\frac{1}{6\sqrt{3}} & \frac{1}{3\sqrt{2}} &
\frac{5}{9\sqrt{10}} \\
&  &  &  &  &  & 2 & -\frac{1}{3\sqrt{6}} & \frac{5}{3\sqrt{30}} \\
&  &  &  &  &  &  & \frac{7}{3} & -\frac{5}{6\sqrt{5}} \\
&  &  &  &  &  &  &  & \frac{19}{9}
\end{array}
\right)
\end{eqnarray*}
\vspace{0.5 cm}

The rank of the idempotent matrices of this matrix is 4. The diagonalization
of the matrix $\left\langle\;H_{\mbox{\small c.m.\vphantom{intr}}%
}\;\right\rangle$ is accomplished within the framework of the outlined above
method using\vspace{0.5 cm}

\begin{eqnarray*}
\left\langle\; a_{K{\mit\Delta};00,{\mit\Gamma}}\;\right\rangle_{\scriptstyle%
 A=3,E=2}^{\scriptstyle J{\mit\Pi}T={\frac{1}{2}}^+{\frac{1}{2}}} =
\left(
\matrix{
\frac{1}{\sqrt{2}}& 0& 0& 0& \cr
-\frac{1}{3\sqrt{2}}&\frac{2}{3}& 0& 0&\cr
0&0&\frac{1}{\sqrt{3}}& 0& \cr 0&-\frac{1}{\sqrt{6}}&0&\frac{1}{3}& \cr
\frac{2}{9\sqrt{3}}&\frac{1}{9\sqrt{6}}&-\frac{2}{9\sqrt{10}}&-\frac{1}{3}&
\cr
\frac{4}{9\sqrt{3}}&\frac{2}{9\sqrt{6}}&\frac{5}{9\sqrt{10}}&\frac{1}{3}&
\cr
-\frac{4}{9}&-\frac{2}{9\sqrt{2}}&-\frac{5}{3\sqrt{30}}&\frac{1}{\sqrt{3}}&
\cr 0&-\frac{1}{\sqrt{3}}&0&-\frac{1}{3\sqrt{2}}& \cr
-\frac{20}{9\sqrt{30}}&-\frac{5}{9\sqrt{15}}&\frac{2}{9}&-\frac{5}{3%
\sqrt{10}}& \cr } \right)
\end{eqnarray*}
\vspace{0.5 cm}

The corresponding CESO's are presented in Table 4.

The intrinsic density matrices in the case of $A = 3,\; E=2,\; JT={\frac{1}{2}%
}{\frac{1}{2}}$ are:

\begin{eqnarray*}
\begin{array}{c}
{\bf W}(\,\raisebox{6pt}{\scriptsize 1}S_0) = \left(
\begin{array}{r}
\frac{3}{8}
\end{array}
\right)
\end{array}
\ ,
\begin{array}{c}
{\bf W}(\,\raisebox{6pt}{\scriptsize 1}S_0^{\prime}) = \left(
\begin{array}{rr}
\frac{1}{24} & -\frac{1}{6\sqrt{2}} \\
& \frac{1}{3}
\end{array}
\right)
\end{array}
\ ,
\begin{array}{c}
{\bf W}(\,\raisebox{6pt}{\scriptsize 3}D_1) = \left(
\begin{array}{r}
\frac{1}{4}
\end{array}
\right)
\end{array}
\ ,
\end{eqnarray*}

\begin{eqnarray*}
\begin{array}{c}
{\bf W}(\,\raisebox{6pt}{\scriptsize 3}S_1) = \left(
\begin{array}{rrr}
\frac{1}{24} & -\frac{1}{6\sqrt{2}} & \hphantom{2}0\hphantom{1} \\
& \frac{1}{3} & \hphantom{2}0\hphantom{1}  \\
&  & \hphantom{2}\frac{1}{4}\hphantom{2}
\end{array}
\right)
\end{array}
\ ,
\begin{array}{c}
{\bf W}(\,\raisebox{6pt}{\scriptsize 3}P_2) = \left(
\begin{array}{rrrr}
\frac{5}{108} & \frac{5}{54\sqrt{2}} & -\frac{5}{36\sqrt{30}} & \frac{5}{36%
\sqrt{3}} \\
& \frac{5}{54} & -\frac{5}{36\sqrt{15}} & \frac{5}{18\sqrt{6}} \\
&  & \frac{1}{72} & -\frac{5}{36\sqrt{10}} \\
&  &  & \frac{5}{36}
\end{array}
\right)
\end{array}
\ ,
\end{eqnarray*}

\begin{eqnarray*}
\begin{array}{c}
{\bf W}(\,\raisebox{6pt}{\scriptsize 3}P_0) = \left(
\begin{array}{rrrr}
\frac{1}{108} & \frac{1}{54\sqrt{2}} & -\frac{5}{18\sqrt{30}} & \frac{1}{18%
\sqrt{3}} \\
& \frac{1}{54} & -\frac{5}{18\sqrt{15}} & -\frac{1}{9\sqrt{6}} \\
&  & \frac{5}{18} & \frac{5}{9\sqrt{10}} \\
&  &  & \frac{1}{9}
\end{array}
\right)
\end{array}
\ ,
\begin{array}{c}
{\bf W}(\,\raisebox{6pt}{\scriptsize 1}P_1) = \left(
\begin{array}{rrr}
\frac{1}{12} & \frac{1}{6\sqrt{2}} & \hphantom{2}0\hphantom{1}  \\
& \frac{1}{6} & \hphantom{2}0\hphantom{1}  \\
&  & \hphantom{2}\frac{1}{2}\hphantom{2}
\end{array}
\right)
\end{array}
\ ,
\end{eqnarray*}

\begin{eqnarray*}
\begin{array}{c}
{\bf W}(\,\raisebox{6pt}{\scriptsize 3}P_1) = \left(
\begin{array}{rrrr}
\frac{1}{36} & \frac{1}{18\sqrt{2}} & \frac{5}{12\sqrt{30}} & -\frac{1}{12%
\sqrt{3}} \\
& \frac{1}{18} & \frac{5}{12\sqrt{15}} & -\frac{1}{6\sqrt{6}} \\
&  & \frac{5}{24} & -\frac{5}{12\sqrt{10}} \\
&  &  & \frac{1}{4}
\end{array}
\right)
\end{array}
\ ,
\begin{array}{c}
{\bf W}(\,\raisebox{6pt}{\scriptsize 3}S_1^{\prime}) = \left(
\begin{array}{r}
\frac{3}{8}
\end{array}
\right)
\end{array}
\ .
\end{eqnarray*}

\noindent Here the primes differentiate the notations of the states with
coinciding orbital momenta.

So far, for all excitation energies under consideration, only the subspaces
corresponding to the nonspurious motion were extracted.
It should be mentioned, that the same procedure can be carried out also for
other subspaces corresponding to the possible excitements of the
centre of mass.
For this, only the value of excitation energy of the centre of mass
(the one from the sequence
${\frac{3}{2}}, {\frac{5}{2}}\ldots E - E_{\min} + {\frac{3}{2}}$)
should be changed.

\section{Conclusions}

The proposed new method for elimination of 'spurious' states enables one to
produce two-particle oscillator intrinsic density matrices without exploiting
the ICFP's which untill now are not available for an arbitrary number of
oscillator quanta.
The distinct advantage of this method is in the complete rejection of
group-theoretical classification of antisymmetric and translationally
invariant many--particle states.
So, a benefit could be gained due to simplicity and comprehensibility of
such kind calculations.
Another distinct feature of the method is that it do not involve any
numerical diagonalization and orthogonalization, i.e. it is stable numericaly.
So, the scale of calculations is determined only by available computation
capabilities.
Software tools based on a root rational fraction expression of numbers for
calculation of CESO's and two-particle oscillator intrinsic density
matrices were developed and implemented in a computer code.
The theoretical formulation have been illustrated by calculation of
two-particle oscillator intrinsic density matrices in the case of
$A=3$, $E=E_{\min },E_{\min }+1,E_{\min }+2$,
and $J{\mit\Pi }T={\frac 12}^{+}{\frac 12}$.

The presented procedure for projecting off the required oscillator shell
model subspace can be generalized by proposing a new method for calculation
of analytical expressions of eigenvectors if the analytical expressions of
eigenvalues and entries of the real symmetric matrix under consideration are
known.
Since it is the case for all quadratic Casimir operators, this method can be
applied to produce large irreducible subspaces with desired transformational
properties.

The new method for elimination of 'spurious' states proposed in this paper
enables one to formulate the problem of calculation of two-particle oscillator
intrinsic density matrices in such form where this problem is no more
complicated as matrix algebra and hence can be solved in a simple and efficient way.

\newpage
Table 1. \vspace{0.5 cm}

The coefficients of expansion of the oscillator shell model functions in terms
of the $A$-particle oscillator functions with singled out dependence on the intrinsic
coordinates of two last particles and with eliminated spurious states
for the three-nucleon system:

\noindent
$E=0, \; K=\left(00{\frac{1}{2}}\right)^3{\frac{1}{2}}{\frac{1}{2}}, \;
\Delta=1, \; J{\mit\Pi}T={\frac{1}{2}}^+{\frac{1}{2}}\ .$ \vspace{0.5 cm}

\begin{tabular}{|c|c|c|c|c|c|c|}
\hline
$\overline{\overline{K\Delta}}$ & $\overline{\overline{JT}}$ & $(el)_{-1}$ &
$elsjt$ & state & $J^{\prime\prime}T^{\prime\prime}$ & CESO's \\ \hline
$\left(00{\frac{1}{2}}\right){\frac{1}{2}}{\frac{1}{2}}$ & ${\frac{1}{2}}{%
\frac{1}{2}}$ & 00 & 00001 & $\raisebox{6pt}{\scriptsize 1}S_0$ & 01 & $-%
\frac{1}{\sqrt{2}}$ \\
&  & 00 & 00110 & $\raisebox{6pt}{\scriptsize 3}S_1$ & 10 & $\frac{1}{\sqrt{2%
}}$ \\ \hline
\end{tabular}
\vspace{1 cm}

Table 2. \vspace{0.5 cm}

The coefficients of expansion of the oscillator shell model functions in terms
of the ones with singled out dependence on the intrinsic coordinates of two
last particles for the three-nucleon system:
$E=1, \; K=\left(00{\frac{1}{2}}\right)^201\left(11{\frac{1}{2}}\right){%
\frac{1}{2}}{\frac{1}{2}}, \; \Delta=1, \; J{\mit\Pi}T={\frac{1}{2}}^-{\frac{%
1}{2}}\ .$ \vspace{0.5 cm}

\begin{tabular}{|c|c|c|c|c|c|c|}
\hline
$\overline{\overline{K\Delta}}$ & $\overline{\overline{JT}}$ & $(el)_{-1}$ &
$elsjt$ & state & $J^{\prime\prime}T^{\prime\prime}$ & SCFP's \\ \hline
$\left(11{\frac{1}{2}}\right){\frac{1}{2}}{\frac{1}{2}}$ & ${\frac{1}{2}}{%
\frac{1}{2}}$ & 00 & 00001 & $\raisebox{6pt}{\scriptsize 1}S_0$ & 01 & $-%
\frac{1}{\sqrt{3}}$ \\[5pt] \hline
$\left(00{\frac{1}{2}}\right){\frac{1}{2}}{\frac{1}{2}}$ &  & 11 & 00110 & $%
\raisebox{6pt}{\scriptsize 3}S_1$ & 00 & $\frac{1}{2\sqrt{2}}$ \\
&  & 00 & 11101 & $\raisebox{6pt}{\scriptsize 3}P_0$ & 01 & $-\frac{1}{2%
\sqrt{6}}$ \\
&  &  & 11010 & $\raisebox{6pt}{\scriptsize 1}P_1$ & 10 & $-\frac{1}{2\sqrt{2%
}}$ \\
&  & 11 & 00110 & $\raisebox{6pt}{\scriptsize 3}S_1$ &  & $\frac{1}{2}$ \\
&  & 00 & 11111 & $\raisebox{6pt}{\scriptsize 3}P_1$ & 11 & $-\frac{1}{2%
\sqrt{3}}$ \\
&  & 11 & 00001 & $\raisebox{6pt}{\scriptsize 1}S_0$ &  & $\frac{1}{2\sqrt{6}%
}$ \\[5pt] \hline
\end{tabular}
\vspace{1 cm}
\newpage

Table 2 (continued). \vspace{0.5 cm}

The coefficients of expansion of the oscillator shell model functions in terms
of the ones with singled out dependence on the intrinsic coordinates of two
last particles for the three-nucleon system:
$E=1, \; K=\left(00{\frac{1}{2}}\right)^210\left(11{\frac{1}{2}}\right){%
\frac{1}{2}}{\frac{1}{2}}, \; \Delta=1, \; J{\mit\Pi}T={\frac{1}{2}}^-{\frac{%
1}{2}}$. \vspace{0.5 cm}

\begin{tabular}{|c|c|c|c|c|c|c|}
\hline
$\overline{\overline{K\Delta}}$ & $\overline{\overline{JT}}$ & $(el)_{-1}$ &
$elsjt$ & state & $J^{\prime\prime}T^{\prime\prime}$ & SCFP's \\ \hline
$\left(11{\frac{1}{2}}\right){\frac{1}{2}}{\frac{1}{2}}$ & ${\frac{1}{2}}{%
\frac{1}{2}}$ & 00 & 00110 & $\raisebox{6pt}{\scriptsize 3}S_1$ & 10 & $-%
\frac{1}{\sqrt{3}}$ \\[5pt] \hline
$\left(00{\frac{1}{2}}\right){\frac{1}{2}}{\frac{1}{2}}$ &  & 11 &  &  & 00
& $\frac{1}{2\sqrt{2}}$ \\
&  & 00 & 11101 & $\raisebox{6pt}{\scriptsize 3}P_0$ & 01 & $\frac{3}{2\sqrt{%
6}}$ \\
&  &  & 11010 & $\raisebox{6pt}{\scriptsize 1}P_1$ & 10 & $\frac{1}{6\sqrt{2}%
}$ \\
&  & 11 & 00110 & $\raisebox{6pt}{\scriptsize 3}S_1$ &  & $-\frac{1}{6}$ \\
&  & 00 & 11111 & $\raisebox{6pt}{\scriptsize 3}P_1$ & 11 & $-\frac{1}{2%
\sqrt{3}}$ \\
&  & 11 & 00001 & $\raisebox{6pt}{\scriptsize 1}S_0$ &  & $\frac{1}{2\sqrt{6}%
}$ \\[5pt] \hline
\end{tabular}
\vspace{1 cm}

Table 2 (continued). \vspace{0.5 cm}

The coefficients of expansion of the oscillator shell model functions in terms
of the ones with singled out dependence on the intrinsic coordinates of two
last particles for the three-nucleon system:
$E=1, \; K=\left(00{\frac{1}{2}}\right)^210\left(11{\frac{3}{2}}\right){%
\frac{3}{2}}{\frac{1}{2}}, \; \Delta=1, \; J{\mit\Pi}T={\frac{1}{2}}^-{\frac{%
1}{2}}$. \vspace{0.5 cm}

\begin{tabular}{|c|c|c|c|c|c|c|}
\hline
$\overline{\overline{K\Delta}}$ & $\overline{\overline{JT}}$ & $(el)_{-1}$ &
$elsjt$ & state & $J^{\prime\prime}T^{\prime\prime}$ & SCFP's \\ \hline
$\left(11{\frac{3}{2}}\right){\frac{3}{2}}{\frac{1}{2}}$ & ${\frac{3}{2}}{%
\frac{1}{2}}$ & 00 & 00110 & $\raisebox{6pt}{\scriptsize 3}S_1$ & 10 & $%
\frac{1}{\sqrt{3}}$ \\[5pt] \hline
$\left(00{\frac{1}{2}}\right){\frac{1}{2}}{\frac{1}{2}}$ & ${\frac{1}{2}}{%
\frac{1}{2}}$ &  & 11010 & $\raisebox{6pt}{\scriptsize 1}P_1$ &  & $\frac{1}{%
3}$ \\
&  & 11 & 00110 & $\raisebox{6pt}{\scriptsize 3}S_1$ &  & $\frac{1}{3\sqrt{2}%
}$ \\
&  & 00 & 11111 & $\raisebox{6pt}{\scriptsize 3}P_1$ & 11 & $\frac{1}{\sqrt{6%
}}$ \\
&  & 11 & 00001 & $\raisebox{6pt}{\scriptsize 1}S_0$ &  & $\frac{1}{\sqrt{3}}
$ \\[5pt] \hline
\end{tabular}
\vspace{1 cm}
\newpage

Table 3. \vspace{0.5 cm}

The coefficients of expansion of the oscillator shell model functions in terms
of $A$-particle oscillator functions with singled out dependence on the intrinsic
coordinates of two last particles and with eliminated spurious states
for the three-nucleon system :

\noindent
$A=3, E=1,\; JT={\frac{1}{2}}{\frac{1}{2}}\ .$ \vspace{0.5 cm}

\begin{tabular}{|c|c|c|c|c|c|c|c|}
\hline
$\overline{\overline{K\Delta}}$ & $\overline{\overline{JT}}$ & $(el)_{-1}$ &
$elsjt$ & state & $J^{\prime\prime}T^{\prime\prime}$ & \multicolumn{2}{|c|}{%
\hphantom{12}CESO's\hphantom{12}} \\ \cline{7-8}
&  &  &  &  &  & $\mit{\Gamma}=1$ & $\mit{\Gamma}=2$ \\ \hline
$\left(00{\frac{1}{2}}\right){\frac{1}{2}}{\frac{1}{2}}$ & ${\frac{1}{2}}{%
\frac{1}{2}}$ & 11 & 00001 & $\raisebox{6pt}{\scriptsize 1}S_0$ & 11 & $-%
\frac{1}{2\sqrt{3}}$ & 0 \\
&  &  & 00110 & $\raisebox{6pt}{\scriptsize 3}S_1$ & 00 & $\frac{1}{6}$ & $%
\frac{1}{3}$ \\
&  &  &  &  & 10 & $\frac{1}{3\sqrt{2}}$ & $-\frac{1}{3\sqrt{2}}$ \\
&  & 00 & 11101 & $\raisebox{6pt}{\scriptsize 3}P_0$ & 01 & $-\frac{1}{2%
\sqrt{3}}$ & $\frac{1}{\sqrt{3}}$ \\
&  &  & 11010 & $\raisebox{6pt}{\scriptsize 1}P_1$ & 10 & $-\frac{1}{2}$ & 0
\\
&  &  & 11111 & $\raisebox{6pt}{\scriptsize 3}P_1$ & 11 & $-\frac{1}{\sqrt{6}%
}$ & $-\frac{1}{\sqrt{6}}$ \\ \hline
$\left(11{\frac{1}{2}}\right){\frac{1}{2}}{\frac{1}{2}}$ &  &  & 00001 & $%
\raisebox{6pt}{\scriptsize 1}S_0$ & 01 & $-\frac{1}{\sqrt{6}}$ & 0 \\
&  &  & 00110 & $\raisebox{6pt}{\scriptsize 3}S_1$ & 10 & $\frac{1}{3\sqrt{6}%
}$ & $-\frac{4}{3\sqrt{6}}$ \\ \hline
$\left(11{\frac{3}{2}}\right){\frac{3}{2}}{\frac{1}{2}}$ & ${\frac{3}{2}}{%
\frac{1}{2}}$ &  &  &  &  & $-\frac{2}{3\sqrt{3}}$ & $-\frac{1}{3\sqrt{3}}$
\\ \hline
\end{tabular}
\vspace{1 cm}
\newpage

Table 4. \vspace{0.5 cm}

The coefficients of expansion of the oscillator shell model functions in trems
of $A$-particle oscillator functions with singled out dependence on the intrinsic
coordinates of two last particles and with eliminated spurious states
for the three-nucleon system :

\noindent
$A=3, E=2,\; JT={\frac{1}{2}}{\frac{1}{2}}\ .$ \vspace{0.5 cm}

\begin{tabular}{|c|c|c|c|c|c|c|c|c|c|}
\hline
$\overline{\overline{K\Delta}}$ & $\overline{\overline{JT}}$ & $(el)_{-1}$ &
$elsjt$ & state & $J^{\prime\prime}T^{\prime\prime}$ & \multicolumn{4}{|c|}{%
\hphantom{12}CESO's\hphantom{12}} \\ \cline{7-10}
&  &  &  &  &  & $\displaystyle\mit{\Gamma} = 1$ & $\mit{\Gamma} = %
\displaystyle 2$ & $\displaystyle\mit{\Gamma} = 3$ & $\mit{\Gamma} = %
\displaystyle 4$ \\ \hline
$\left(00{\frac{1}{2}}\right){\frac{1}{2}}{\frac{1}{2}}$ & ${\frac{1}{2}}{%
\frac{1}{2}}$ & 20 & 00001 & $\raisebox{6pt}{\scriptsize 1}S_0$ & 01 & $-%
\frac{1}{2\sqrt{6}}$ &  0&  0& 0 \\
&  & 00 & 20001 &  &  & $-\frac{1}{2\sqrt{6}}$ & $\frac{1}{\sqrt{3}}$ & 0 & 0
\\
&  & 20 & 00110 & $\raisebox{6pt}{\scriptsize 3}S_1$ & 10 & $\frac{1}{6\sqrt{%
6}}$ & $-\frac{1}{3\sqrt{3}}$ & 0 & 0 \\
&  & 22 &  &  &  & 0 & 0 & $-\frac{1}{6}$ & 0 \\
&  & 00 & 20110 &  &  & $\frac{3}{2\sqrt{6}}$ & 0 & 0 & 0 \\
&  & 11 & 11101 & $\raisebox{6pt}{\scriptsize 3}P_0$ & 11 & $-\frac{1}{18}$
& $-\frac{1}{9\sqrt{2}}$ & $\frac{5}{3\sqrt{30}}$ & $\frac{1}{3\sqrt{3}}$ \\
&  &  & 11010 & $\raisebox{6pt}{\scriptsize 1}P_1$ & 00 & $\frac{1}{6}$ & $%
\frac{1}{3\sqrt{2}}$ & 0 & 0 \\
&  &  &  &  & 10 & 0 & 0 & 0 & $\frac{1}{\sqrt{6}}$ \\
&  &  & 11111 & $\raisebox{6pt}{\scriptsize 3}P_1$ & 01 & 0 & 0 & 0 &
$-\frac{1}{3\sqrt{2}}$ \\
&  &  &  &  & 11 & $\frac{1}{6\sqrt{3}}$ & $\frac{1}{3\sqrt{6}}$ & $\frac{5}{%
6\sqrt{10}}$ & $-\frac{1}{6}$ \\
&  &  & 11121 & $\raisebox{6pt}{\scriptsize 3}P_2$ &  & $-\frac{5}{18\sqrt{5}%
}$ & $-\frac{5}{9\sqrt{10}}$ & $\frac{1}{6\sqrt{6}}$ & $-\frac{5}{6\sqrt{15}}
$ \\
&  & 00 & 22110 & $\raisebox{6pt}{\scriptsize 3}D_1$ & 10 & 0 & 0 &
$\frac{1}{2}$ & 0 \\[5pt] \hline
$\left(11{\frac{1}{2}}\right){\frac{1}{2}}{\frac{1}{2}}$ &  & 11 & 00001 & $%
\raisebox{6pt}{\scriptsize 1}S_0$ & 11 & $\frac{1}{3\sqrt{2}}$ & 0 & 0 & 0 \\
&  &  & 00110 & $\raisebox{6pt}{\scriptsize 3}S_1$ & 00 & $\frac{1}{9\sqrt{6}%
}$ & $-\frac{2}{9\sqrt{3}}$ & $-\frac{5}{9\sqrt{5}}$ & 0 \\
&  &  &  &  & 10 & $-\frac{1}{9\sqrt{3}}$ & $\frac{4}{9\sqrt{6}}$ & $-\frac{5%
}{9\sqrt{10}}$ & 0 \\
&  & 00 & 11101 & $\raisebox{6pt}{\scriptsize 3}P_0$ & 01 & $-\frac{1}{9%
\sqrt{2}}$ & $-\frac{1}{9}$ & $\frac{5}{3\sqrt{15}}$ & $\frac{2}{3\sqrt{6}}$
\\
&  &  & 11010 & $\raisebox{6pt}{\scriptsize 1}P_1$ & 10 & $-\frac{1}{3\sqrt{6%
}}$ & $-\frac{1}{3\sqrt{3}}$ & 0 & $\frac{2}{3\sqrt{2}}$ \\
&  &  & 11111 & $\raisebox{6pt}{\scriptsize 3}P_1$ & 11 & $\frac{1}{9}$ & $%
\frac{2}{9\sqrt{2}}$ & $\frac{5}{3\sqrt{30}}$ & 0 \\[5pt] \hline
\end{tabular}

\newpage

Table 4 (continued). \vspace{0.5 cm}

The coefficients of expansion of the oscillator shell model functions in trems
of $A$-particle oscillator functions with singled out dependence on the intrinsic
coordinates of two last particles and with eliminated spurious states
for the three-nucleon system :

\noindent
$A=3, E=2,\; JT={\frac{1}{2}}{\frac{1}{2}}\ .$ \vspace{0.5 cm}

\begin{tabular}{|c|c|c|c|c|c|c|c|c|c|}
\hline
$\overline{\overline{K\Delta}}$ & $\overline{\overline{JT}}$ & $(el)_{-1}$ &
$elsjt$ & state & $J^{\prime\prime}T^{\prime\prime}$ & \multicolumn{4}{|c|}{%
\hphantom{12}CESO's\hphantom{12}} \\ \cline{7-10}
&  &  &  &  &  & $\mit{\Gamma}=1$ & $\mit{\Gamma}=2$ & $\mit{\Gamma}=3$ & $%
\mit{\Gamma}=4$ \\ \hline
$\left(11{\frac{3}{2}}\right){\frac{3}{2}}{\frac{1}{2}}$ & ${\frac{3}{2}}{%
\frac{1}{2}}$ & 11 & 00001 & $\raisebox{6pt}{\scriptsize 1}S_0$ & 11 & $%
\frac{1}{3}$ & 0 & 0 & 0 \\
&  &  & 00110 & $\raisebox{6pt}{\scriptsize 3}S_1$ & 10 & $\frac{1}{9\sqrt{6}%
}$ & $-\frac{2}{9\sqrt{3}}$ & $\frac{5}{18\sqrt{5}}$ & 0 \\
&  &  &  &  & 20 & $-\frac{5}{9\sqrt{30}}$ & $\frac{10}{9\sqrt{15}}$ & $%
\frac{1}{18}$ & 0 \\
&  & 00 & 11010 & $\raisebox{6pt}{\scriptsize 1}P_1$ & 10 & $-\frac{1}{3%
\sqrt{3}}$ & $-\frac{2}{3\sqrt{6}}$ & 0 & $-\frac{1}{3}$ \\
&  &  & 11111 & $\raisebox{6pt}{\scriptsize 3}P_1$ & 11 & $-\frac{1}{9\sqrt{2%
}}$ & $-\frac{1}{9}$ & $-\frac{5}{6\sqrt{15}}$ & $\frac{1}{\sqrt{6}}$ \\
&  &  & 11121 & $\raisebox{6pt}{\scriptsize 3}P_2$ & 21 & $\frac{5}{9\sqrt{10%
}}$ & $\frac{5}{9\sqrt{5}}$ & $-\frac{1}{6\sqrt{3}}$ & $\frac{5}{3\sqrt{30}}$
\\[5pt] \hline
$\left(20{\frac{1}{2}}\right){\frac{1}{2}}{\frac{1}{2}}$ & ${\frac{1}{2}}{%
\frac{1}{2}}$ &  & 00001 & $\raisebox{6pt}{\scriptsize 1}S_0$ & 01 &
$-\frac{1}{\sqrt{6}}$ & 0 & 0 & 0 \\
&  &  & 00110 & $\raisebox{6pt}{\scriptsize 3}S_1$ & 10 & $\frac{1}{3\sqrt{6}%
}$ & $-\frac{2}{3\sqrt{3}}$ & 0 & 0 \\[5pt] \hline
$\left(22{\frac{3}{2}}\right){\frac{3}{2}}{\frac{1}{2}}$ & ${\frac{3}{2}}{%
\frac{1}{2}}$ &  &  &  &  & 0 & 0 & $\frac{1}{3}$ & 0 \\[5pt] \hline
\end{tabular}
\vspace{1 cm}

\end{document}